\documentclass[preprint,aps,12pt,showpacs,nofootinbib,tightenlines,amsmath,amssymb]{revtex4}
\usepackage{stmaryrd}

\usepackage{amsmath}
\usepackage{graphicx}
\usepackage{amssymb}

\begin{document}
\def\intdk{\int\frac{d^4k}{(2\pi)^4}}
\def\sla{\hspace{-0.17cm}\slash}
\hfill

\title{Entropy Bound with Generalized Uncertainty Principle in General Dimensions }

\author{Weijian Wang}\email{wjnwang96@gmail.com}\affiliation{Department of Physics, Zhejiang University, Hangzhou
310027, China}
\author{Da Huang} \email{dahuang@itp.ac.cn}\affiliation{Kavli Institute for
Theoretical Physics China (KITPC)
\\ State Key Laboratory of Theoretical Physics (SKLTP) \\ Institute of
Theoretical Physics, Chinese Academy of Science, Beijing,100190,
P.R.China}
\date{\today}

\begin{abstract}
In this letter, the entropy bound for local quantum field theories
(LQFT) is studies in a class of models of the generalized
uncertainty principle(GUP) which predicts a minimal length as a
reflection of the quantum gravity effects. Both bosonic and
fermionic fields confined in arbitrary spatial dimension $d\geq4$
ball ${\cal B}^{d}$ are investigated. It is found that the GUP leads
to the same scaling $A_{d-2}^{(d-3)/(d-2)}$ correction to the
entropy bound for bosons and fermions, although the coefficients of
this correction are different for each case. Based on our
calculation, we conclude that the GUP effects can become manifest at
the short distance scale. Some further implications and speculations
of our results are also discussed.

\end{abstract}
\pacs{11.10.Kk,04.70.Dy,04.50.-h,04.60.Bc,03.67.-a}

\maketitle

\section{Introduction}
Although a full description of quantum gravity is still missing, the
existence of a minimal length appears as a common feature in various
candidates of the quantum gravity, such as string
theory\cite{Witten96,Connes} and black hole physics\cite{Mag1993}.
Motivated by this progress, the idea of minimal length has attracted
great interest in recent years.

Phenomenologically, the existence of a minimal length implies that a
generalized uncertainty principle(GUP) should replace the ordinary
one in the usual quantum mechanics. There are already many different
proposals on the realization of this idea in the
literature\cite{Kem1995}. Recently, a new class of GUP models was
proposed by Ali, Das, and Vagenas(ADV)\cite{Ali} in order to
incorporate the string theory, black hole physics and double special
relativity (DSR). In their proposal, the minimal length can be
characterized by the position and momentum operators satisfying
generalized commutation relations\cite{Ali}:
\begin{equation}\begin{split}
&[x_{i}, x_{j}]=[p_{i}, p_{j}]=0,\\
[x_{i},
p_{j}]=i\hbar[\delta_{ij}&-\alpha(p\delta_{ij}+\frac{p_{i}p_{j}}{p})+\alpha^{2}(p^{2}\delta_{ij}+3p_{i}p_{j})],\\
\label{uncertain}\end{split}\end{equation} which yields the
generalized uncertainty principle to $\alpha^{2}$ order
\begin{equation}
\Delta x\Delta p\geq\frac{\hbar}{2}[1-2\alpha<p>+4\alpha^{2}<p^{2}>]
\end{equation}
where $\alpha=\alpha_{0}/M_{Pl}c=\alpha_{0}l_{Pl}/\hbar$, with
$M_{Pl}$ the Plank mass and $l_{Pl}$ the Plank length. One can check
that Eq.\eqref{uncertain} given above is covariant under DSR
transformation\cite{Cor} which preserves a invariant energy scale.
Thus it is natural to derive both a minimal measurable length and a
maximal measurable momentum.i.e
\begin{equation}\begin{split}
&\Delta x\geq(\Delta x)_{mim}\approx\alpha_{0}l_{Pl},\\
&\Delta p\leq(\Delta p)_{max}\approx\frac{M_{Pl}c}{\alpha_{0}}.\\
\end{split}\end{equation}
Some phenomenological implications of above GUP have been studied in
\cite{Bas,AF}. An interesting effect is the modification of
invariant weighted phase space volume. By rigorously solving the
Liouville problem, the number of quantum states per momentum space
volume is given by\cite{AF}
\begin{equation}
\frac{V}{(2\pi\hbar)^{D}}\frac{d^{D}p}{(1-\alpha p)^{D+1}}.
\label{phase}\end{equation}

On the other hand, it is generally believed that, at the fundamental
level, the "holographic information bound" is an essential aspect of
the nature of the quantum gravity\cite{'tHooft:1993gx,Susskind}. It
states that the maximal entropy $S_{max}$ in a volume $R$ is bounded
by the area of the boundary of the system $A(\partial R)$. A more
stringent constraint can be obtained if one requires that the system
can be described by some local quantum fields theory (LQFT). Cohen
et.al. \cite{Cohen} proposed a entropy bound $S_{max}\sim
A^{\frac{3}{4}}$ for ordinary four-dimension spacetime LQFT by
imposing the gravitational stability condition on the system. The
mismatch of the entropy bound $A^{\frac{3}{4}}$ and holographic
bound $A$ can be understood by the fact that the usual description
of LQFT cannot work in the strongly gravitational circumstance.

Thus, it is interesting to consider the impact of GUP on the entropy
bound, which can be regarded, at least partially, as a reflection of
quantum gravity effect. With the method presented in
\cite{Yurtsever,Aste,Chen}, this investigation\cite{Yong,AF} has
been done for the various GUP models, but their discussion was only
restricted for bosonic fields in four dimension spacetime. In this
letter, we present the entropy bound corrected by the model of
GUP\cite{Ali} whose commutation relations are given in
Eq.\eqref{uncertain}. We do our calculation in the arbitrary
dimension $d\geq4$ and both the bosonic and fermionic field are
considered, generalizing the results given in \cite{AF}. It is found
that this GUP model leads to the same scaling correction
$A_{d-2}^{(d-3)/(d-2)}$ to the ordinary entropy bound
$A_{d-2}^{(d-1)/d}$ for both the bosonic and fermionic systems,
although the concrete coefficients before this correction are quite
different. Based on our calculation, we conclude that the GUP
effects can become manifest at the short distance scale. Some
further implications and speculations of our results are also
discussed.

In this paper, we will use the unit such that $c=\hbar=G=k_B=1$

\section{Derivation of Holographic Entropy Bound for a Real Scalar Boson in GUP Theory}
In this section we give the detailed derivation of holographic
entropy bound for the theory of the generalized uncertainty
principle (GUP)\cite{Ali} in arbitrary spacetime dimension $d$.
First we consider the case of a free massless real scalar field
$\Phi(x)$, which is confined in a $(d-1)$-dimensional spacelike ball
${\cal B}^R_{d-1}$ with the radius $R$, as has been done in Ref.
\cite{Yong, Aste, Chen}. The modes of the field are then the
solutions of the scalar wave equation $\boxempty \Phi=0$ that vanish
on the surface of the ball $\partial{\cal B}^R_{d-1}$, i.e. on the
sphere ${\cal S}^R_{d-2}$. Using \eqref{phase} and taking the
continuous limit, we can perform the summation of quantities over
the modes by the following replacement
\begin{equation}
\sum_{\textbf{p}} \to
\int\frac{d^{d-1}\textbf{x}~d^{d-1}\textbf{p}}{(2\pi)^{d-1}}\frac{1}{(1-\alpha
p)^{d}} = \frac{A_{d-2}(1)V_{d-1}(R)}{(2\pi)^{d-1}}\int
\frac{dp~p^{d-2}}{(1-\alpha p)^d},
\end{equation}
where $V_{d-1}(R)$ is the $(d-1)$-dimensional volume of ${\cal
B}_{d-1}^R$ and $A_{d-2}(R)$ the $(d-2)$-dimensional volume of the
sphere ${\cal S}^R_{d-2}$, which can be explicitly given by:
\begin{equation}
V_n(R) = \frac{2 \pi^{n/2}}{n\Gamma(n/2)}R^n, \quad A_{n-1}(R)=
\frac{2\pi^{n/2}}{\Gamma(n/2)}R^{n-1}.
\end{equation}

In order to derive the holographic entropy bound for the GUP theory,
we need to impose now two {\em ad hoc} restrictions on the
admissible states in the bosonic Fock space. Firstly, we assume that
the maximum energy of the allowed modes is given by the UV cutoff
$\Lambda$ of the local quantum field theory (LQFT). This condition
effectively makes the total number of the quantized modes $N$
finite, which is given by
\begin{eqnarray}\label{mode num}
N &=& \sum_{\textbf{p}} 1 \to
\frac{A_{d-2}(1)V_{d-1}(R)}{(2\pi)^{d-1}}\int^\Lambda_0
\frac{dp~p^{d-2}}{(1-\alpha p)^d}\nonumber\\ &\approx&
\frac{A_{d-2}(1)V_{d-1}(1)}{(2\pi)^{d-1}(d-1)}R^{d-1}\Lambda^{d-1}[1
+\alpha(d-1)\Lambda],
\end{eqnarray}
where we assume that $\alpha<\frac{1}{\Lambda}$ so that we can
approximate the result by expanding the integrand to leading order
in $\alpha$. Note that the total number of allowed modes increases
due to the GUP correction.

Secondly, in order that the configuration of the states would not
collapse to a black hole and the field states to be observable for
outside world\cite{'tHooft:1993gx}, we need to impose the maximum
energy to a state in Fock space which should not be larger than the
mass of the corresponding d-dimensional Schwarzschild black hole
within the same size:
\begin{eqnarray}
E_{BH}(R)= \frac{(d-2)A_{d-2}(R)}{16\pi R} = \eta R^{d-3},
\end{eqnarray}
where, for later convenience, we have defined a dimensionless order
1 parameter $\eta$:
\begin{equation}
\eta \equiv \frac{(d-2)A_{d-2}(1)}{16\pi}.
\end{equation}
 The Fock states can be constructed by assigning
the occupying number $n_i$ to these $N$ different modes,
\begin{eqnarray}
|\Psi> = |n(\textbf{p}_1),n(\textbf{p}_2),...,n(\textbf{p}_N)>\to
|n_1,n_2,...,n_N>.
\end{eqnarray}
The dimension of the Hilbert space is calculated by counting the
possible distributions of the number occupancies $\{n_i\}$. The
non-gravitational-collapse condition leads to finiteness of the
Hilbert space:
\begin{equation}\label{stable}
E = \sum^N_{i=1} n_i p_i \leqslant E_{BH}(R)= \eta R^{d-3}.
\end{equation}
Let us now consider a typical $N$-particle state with one particle
occupying each mode ($n_i=1$), corresponding to the lowest energy
state with $N$ modes simultaneously excited. For this state, the
total energy is
\begin{eqnarray}
E_{tot} &=& \sum_{i=1}^N p_i \to \frac{A_{d-2}(1)V_{d-1}(1)
R^{d-1}}{(2\pi)^{d-1}} \int^\Lambda_0 \frac{dp~p^{d-1}}{(1-\alpha
p)^{d}}\nonumber\\
&\approx& \mu R^{d-1} \Lambda^d [1+\frac{\alpha d^2}{d+1}\Lambda],
\end{eqnarray}
where we also expand the result to the leading order in the small
parameter $\alpha$ and absorb the various irrelevant dimensionless
order 1 coefficients into a single one $\mu$:
\begin{equation}
\mu= \frac{A_{d-2}(1)V_{d-1}(1)}{(2\pi)^{d-1}d}.
\end{equation}
By imposing the gravitational stability condition Eq.(\ref{stable}),
one can arrive at the deformed UV-IR condition \cite{Cohen} due to
the GUP effects
\begin{eqnarray}
R^2 \Lambda^d(1+\frac{\alpha d^2}{d+1}\Lambda)\leqslant
\frac{\eta}{\mu},
\end{eqnarray}
or, inversely, we can rewrite this condition in another form which
is more convenient to use in the following:
\begin{eqnarray}\label{UV-IR}
\Lambda^d \leqslant \frac{\eta}{\mu}\frac{1}{R^2}[1-\frac{\alpha
d^2}{d+1}(\frac{\eta}{\mu})^{1/d}R^{-2/d}].
\end{eqnarray}

The entropy associated with the system is
\begin{equation}
S= -\sum^{W^b}_{i=1}\rho_i \ln\rho_i,
\end{equation}
where $\rho_i$ is the possibility distribution on the Hilbert state
basis and $W^b$ is the dimension of Hilbert space $W^b=
{\text{dim}}{\cal H}$. The maximum value of the expression can be
reached by a uniform distribution
$\rho_i=\frac{1}{W}$\cite{Yurtsever,Aste,Yong,Chen}. So the maximum
entropy is given by
\begin{equation}
S_{max} = \ln W^b.
\end{equation}
The bound of the dimension of the Hilbert space is studied in
\cite{Yurtsever,Aste,Chen} and is determined by following formula:
\begin{equation}
W^b=\text{dim}{\cal H} < \sum^N_{m=0}\frac{z^m}{(m!)^2} \leqslant
\sum^\infty_{m=0}\frac{z^m}{(m!)^2} = I_0(2\sqrt{z})\sim
\frac{e^{2\sqrt{z}}}{\sqrt{4\pi\sqrt{z}}},
\end{equation}
where $I_0(x)$ is the zeroth-order Bessel function of the second
kind. And $z$ is given by
\begin{eqnarray}
z &=& \sum^N_{i=1} \frac{E_{BH}(R)}{p} \to
\frac{A_{d-2}(1)V_{d-1}(R)}{(2\pi)^{d-1}}\int^\Lambda_0
\frac{E_{BH}(R)}{p}\frac{p^{d-2}d p}{(1-\alpha p)^d}\nonumber\\
&\approx& \mu\eta \frac{d}{d-2}
R^{2(d-2)}\Lambda^{d-2}(1+\frac{\alpha d(d-2)}{d-1}\Lambda).
\end{eqnarray}
With the gravitational stability condition Eq.(\ref{stable}), we can
further obtain the bound written in terms of the radius $R$ which
can also be regarded as the IR cutoff:
\begin{eqnarray}
z \leqslant \mu\eta\frac{d}{d-2}(\frac{\eta}{\mu})^{(d-2)/d}
R^{2(d-1)(d-2)/d}[1+2\alpha\frac{
d(d-2)}{(d+1)(d-1)}(\frac{\eta}{\mu})^{1/d}R^{-2/d}].
\end{eqnarray}
So the maximum entropy for the scalar field in the GUP background is
\begin{eqnarray}\label{boson}
S_{max}^b &=& \ln W^b \approx 2\sqrt{z}\nonumber\\
 &\leqslant& 2
 \sqrt{\mu\eta\frac{d}{d-2}}~(\frac{\eta}{\mu})^{\frac{d-2}{2d}}
 R^{(d-1)(d-2)/d}[1+\frac{\alpha
 d(d-2)}{(d+1)(d-1)}(\frac{\eta}{\mu})^{1/d}R^{-2/d}]\nonumber\\
&\propto& A_{d-2}^{(d-1)/d}+ \alpha\frac{
d(d-2)}{(d+1)(d-1)}(\frac{\eta}{\mu})^{1/d} A_{d-2}^{(d-3)/(d-2)},
\end{eqnarray}
where $A_{d-2}$ is the area of the sphere enclosing the ball ${\cal
B}^R_{d-1}$.

A simple but perhaps the most important example of this formula is
the case when d=4, the holographic entropy bound for the scalar
field is:
\begin{eqnarray}
S_{max}^b(d=4) \leqslant A_2^{3/4} + \frac{16}{30}\alpha
(\frac{\eta}{\mu})^{1/4} A_2^{1/2}
\end{eqnarray}
which agrees with the result given in \cite{AF}

\section{Derivation of Holographic Entropy Bound for a Fermion in GUP Theory}
This section is devoted to the calculation of the holographic
entropy bound for a fermionic degree of freedom. As the previous
section, we also calculate the entropy bound for the fermion within
a $(d-1)$-dimensional ball. Due to the two constraints introduced in
the previous section, the total number of allowed excited modes and
the UV-IR relation are the same as those in the real scalar case,
which is given in Eq.(\ref{mode num}) and Eq.(\ref{UV-IR})
respectively. However, because of the difference in the statistics
of bosons from fermions, the dimension of the Hilbert space is not
the same as before and we need some new insights for this case.

In fact, the counting of dimension of the Hilbert space is quite
simple for the fermion case. By taking into account the Pauli
exclusion principle, every mode can only have two states: empty and
occupying. So for effective N excited modes, there are $2^N$ states.
Thus, the dimension of the Hilbert space is:
\begin{equation}
W^f = 2^N.
\end{equation}

Following the arguments before, the distribution with the maximum
entropy is the uniform one, so the maximum entropy is given by:
\begin{equation}
S_{max} = \ln W = \ln2\cdot N.
\end{equation}
Due to the inequality Eq.(\ref{UV-IR}) given by the gravitational
stability, we can obtain a maximum bound for the allowed number of
modes $N$, and in turn the bound for the entropy:
\begin{eqnarray}
N &\approx&
\frac{A_{d-2}(1)V_{d-1}(1)}{(2\pi)^{d-1}(d-1)}R^{d-1}\Lambda^{d-1}[1
+\alpha(d-1)\Lambda]\nonumber\\
&\leqslant& \mu\frac{d}{d-1} (\frac{\eta}{\mu})^\frac{d-1}{d}
R^{(d-1)(d-2)/d}[1+\alpha\frac{d-1}{d+1}(\frac{\eta}{\mu})^{1/d}R^{-2/d}].
\end{eqnarray}
Therefore the holographic bound for a fermionic degree of freedom
with the GUP effects is given by
\begin{equation}\label{ferm}
S_{max}^f \leqslant A_{d-2}^{(d-1)/d} +\alpha
\frac{d-1}{d+1}(\frac{\eta}{\mu})^{1/d}A^{(d-3)/(d-2)}_{d-2}.
\end{equation}
For the case of dimension 4 spacetime, we have
\begin{equation}
S_{max}^f \leqslant A_{2}^{3/4} + \frac{3}{5}
\alpha(\frac{\eta}{\mu})^{1/4}A^{1/2}_{2}.
\end{equation}
Note that both the bosonic and fermionic fields share the same
$A_{d-2}^{(d-3)/(d-2)}$ scaling correction to the entropy bound due
to the GUP, even though the coefficients before this correction are
different for the two cases.

\section{Conclusion and Discussion}
In this paper, we study the impact of generalized uncertainty
principle(GUP) proposed recently in \cite{Ali} on the entropy bound
for a local quantum field theory (LQFT) system in arbitrary
dimensions, which can be regarded as the effects of quantum gravity.
Both the bosonic and fermionic fields are considered. We find that
the leading order $\alpha$ correction to the entropy bound goes as a
positive one $A_{d-2}^{(d-3)/(d-2)}$ for both cases, although the
coefficients before this correction are quite different. According
to our calculations above, we know that when the volume of ${\cal
S}^{d-2}$ is very large, the leading term $A_{d-2}^{(d-1)/d}$ in
Eqs.(\ref{boson}) and (\ref{ferm}) dominates and the correction due
to GUP effects can be ignorable. However, if volume of ${\cal
S}^{d-2}$ becomes smaller and reach the scale $A_{d-2}\sim
\alpha^{d(d-2)/2}$, the GUP correction can compete with the ordinary
leading term. When the volume becomes even smaller and approaches
zero, the GUP effects begin to dominates the scaling behavior.
Although the picture discussed above is only based upon our small
$\alpha$ approximation and higher order terms in the expansion of
$\alpha$ may change the scaling behavior dramatically, it is still
sufficient to conclude that if the GUP really exists in nature, it
can only manifest itself at the very short distance scale. This
conclusion agrees with our usual straightforward intuition and the
explanation that the GUP is the manifestation of the quantum gravity
at some short distance scale.

Our calculation in this paper confirms the arguments in \cite{AF}
that the holographic theories lose its good features due to the
violation of continuous symmetry by the discrete space. Moreover, an
interesting prospect comes up if one compares our results with a
heuristic analysis\cite{Maj} of black hole entropy corrected by the
GUP model we considered here. The corrected entropy for
Schwarzschild black hole in four dimension up to the first order is
given by\cite{Maj}
\begin{equation}
S\simeq A+\frac{\sqrt{\pi}}{2}\alpha\sqrt{A}
\end{equation}
Although the leading order is different for black hole entropy bound
and LQFT entropy bound, the scaling behavior of the first-order
correction by the GUP is precisely the same, at least in four
spacetime dimensions. Is it possible that the quantum gravity
effects lead to some universal contributions for both holographic
systems and LQFT systems? This needs more studies for the black hole
entropy with the GUP, especially in general dimensions. The relevant
research is being investigated.

\vspace{1 cm}

\centerline{{\bf Acknowledgement}}

\vspace{20 pt}

\noindent We would like to thank the $1^{\text{st}}$ IAS-CERN School
on Particle Physics and Cosmology and Implications for Technology at
Nanyang Technological University for their hospitality and an
stimulating environment. DH was supported in part by the National
Science Foundation of China (NSFC) under Grant \#No. 10821504,
10975170 and the Project of Knowledge Innovation Program (PKIP) of
the Chinese Academy of Science. W. Wang was supported in part by
funds from NSFC under Grant \#No. 11075140 and Fundamental Research
Funds for the Central University.

\end{document}